\documentclass[twocolumn,prl]{revtex4}
\usepackage{graphicx}
\usepackage{subfigure}
\usepackage{bm}
\usepackage{amssymb}
\usepackage{wasysym}

\newcommand{\Ket}[1]{\vert \, #1 \, \rangle}
\newcommand{\Bra}[1]{\langle \, #1 \,\vert}

\newcommand{\Amp}[2]{\langle \, #1\, \vert\,  #2 \, \rangle}

\newcommand{\Avg}[1]{\langle  #1  \rangle}

\renewcommand{\phi}{\varphi}
\renewcommand{\epsilon}{\varepsilon}

\begin{document}

\title{Topological Transition in a Non-Hermitian Quantum Walk}
\author{M. S. Rudner$^1$ and L. S. Levitov$^{1,2}$}
\affiliation{
$^{(1)}$ Department of Physics,
 Massachusetts Institute of Technology, 77 Massachusetts Ave,
 Cambridge, MA 02139\\
$^{(2)}$ Kavli Institute for Theoretical Physics, University of California, Santa Barbara, CA 93106}

\begin{abstract}
We analyze a quantum walk on a bipartite one-dimensional lattice, in which the particle can decay whenever it visits 
one of the two sublattices.
The corresponding non-Hermitian tight-binding problem with a complex 
potential for the decaying sites exhibits two different phases, distinguished by 
a winding number defined in terms of the Bloch eigenstates in the Brillouin zone.
We find that the mean displacement of a particle initially localized on one of the non-decaying sites can be expressed in terms of the winding number, and is therefore quantized as an integer, changing from zero to one at the critical point. 
This problem can serve as a simplified model for 
nuclear spin pumping in the spin-blockaded electron transport regime of quantum dots in the presence of competing 
hyperfine and spin-orbital interactions. 
The predicted transition from pumping to non-pumping is topological in nature, and is hence robust against certain types of noise and decoherence. 
\end{abstract}

\maketitle


A quantum system is said to exhibit 
a topological transition when it features several 
phases, characterized 
by a 
topological invariant that takes on different quantized values in each of these phases\,\cite{Fradkin}.
A classic example of a topological transition occurs in the quantized Hall effect, which can be linked to the Chern invariant~\cite{HallConductance}, defined in terms of the system's single-particle wavefunctions with quasi-periodic boundary conditions. 
Because the Hall conductance is proportional to 
the Chern invariant, and because quantization of the latter is of a topological nature, 
the quantized Hall effect is universal across samples of varying size, shape, or composition, and is robust against many types of disorder. 
Another example of a similar nature is encountered in adiabatic transport~\cite{AdiabaticTransport}.

\begin{figure}
\includegraphics[width=3.2in]{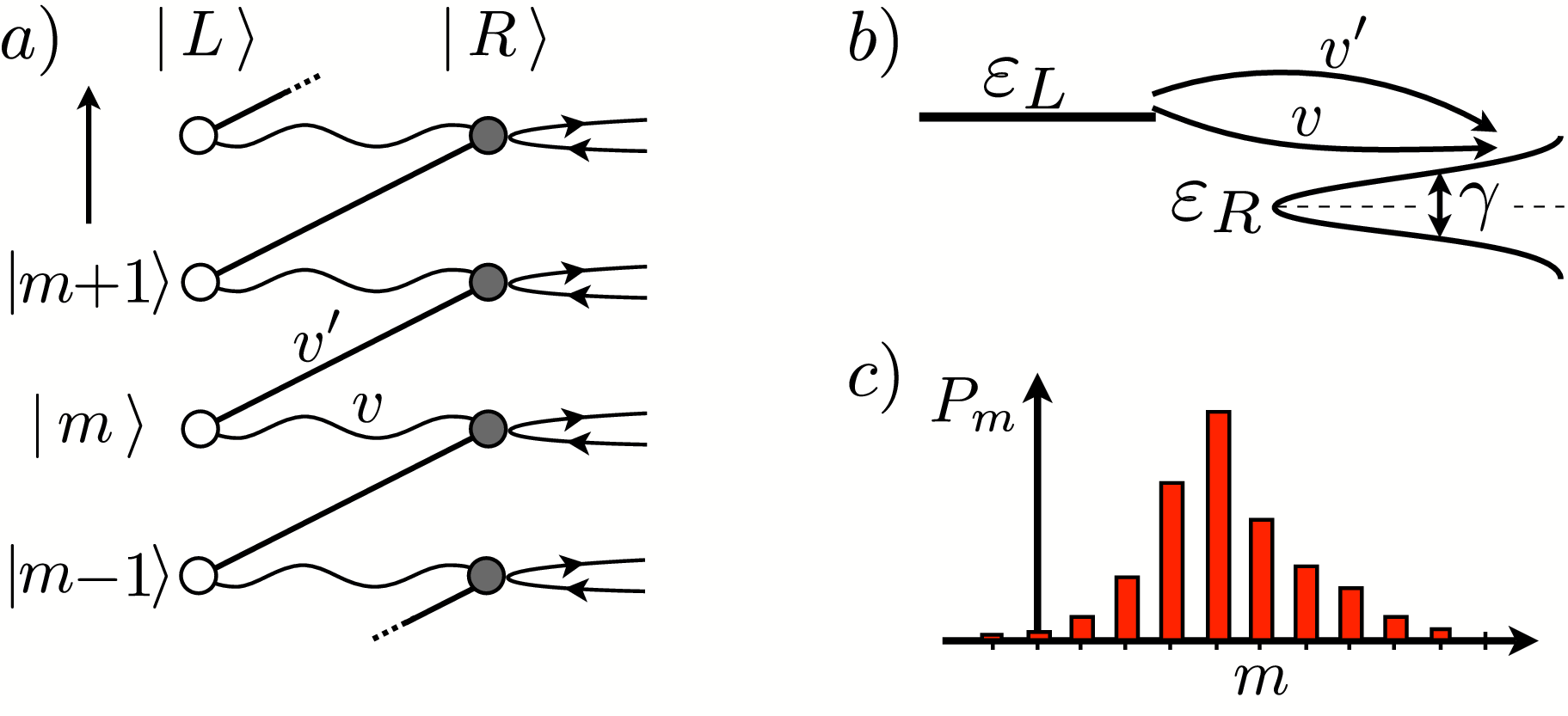}
 \caption[]{Setup of the model.
a) 
Each unit cell $m$ contains two sites $L$ (open circles) and $R$ (filled circles), with each $R$-site connected to an external decay channel.
Intracell (wavy lines) and intercell (straight lines) tunneling occur with amplitudes ${v}$ and ${v'}$, respectively.
b) Energies of the $L$ and $R$ sites. 
Due to decay, 
the $R$-site energy obtains an imaginary part, $\tilde{\varepsilon}_R = \varepsilon_R - i\hbar\gamma/2$. 
c) Schematic distribution of local decay probabilities $\{P_m\}$ used to calculate the displacement (\ref{DeltaM}).
 }
\label{Cartoons}
\vspace{-4mm}
\end{figure}

Here we present a model exhibiting a new type of topological transition in a system described by a non-Hermitian Hamiltonian. 
We consider a quantum walk on a bipartite one-dimensional (1D) lattice, from which the ``walker'' (particle) can decay whenever it resides on the sites of one of the sublattices (see Fig.\ref{Cartoons}a).
Due to hopping between sites, a particle initially localized on any of the non-decaying sites at time $t = 0$ will eventually decay from the system as $t \rightarrow \infty$. 

Surprisingly, we find that the average displacement of the particle during the course of its decay, $\Delta m=\sum_m mP_m$, is exactly quantized as an integer (0 or 1 unit cells), where $P_m$ is the probability distribution for decay from different sites (see Fig.\ref{Cartoons}c).
As in the case of the quantum Hall conductance, this quantization results from an underlying topological structure; in this case it is the winding number of the relative phase between two components of the Bloch wave function.
Using the topological origin of this phenomenon, we are able to show that the quantization is insensitive to parameters and is robust against certain types of noise and decoherence.

In recent years, non-Hermitian quantum mechanics in one dimension has found applications to a variety of different problems, such as vortex matter\cite{Hatano96}, quantum chaos\cite{Efetov97}, random lasers\cite{Beenakker96}, population biology\cite{Shnerb98}, and others (see Ref.\cite{Feinberg97} and references therein). Much of the interest in these 1D problems was triggered by the idea that an Anderson localization transition can occur in disordered transport with an imaginary vector potential\cite{Hatano96}. 
In contrast, our problem is translationally invariant; the transition results from competition between two processes, intracell and intercell hopping, which occur with amplitudes $v$ and $v'$ (see Fig.\ref{Cartoons}a).

In this paper, our motivation is to provide a simple model of nuclear spin pumping in spin-blockaded double quantum dots~\cite{OnoTarucha, Koppens, Pfund} in the presence of competing effects of the hyperfine and spin-orbital interactions, as in Ref.\cite{Pfund}. 
In the DC transport regime, an electron, first loaded into a triplet spin state $\Ket{L}$, makes a transition to a singlet spin state $\Ket{R}$, which is broadened due to its coupling to the drain lead \cite{RudnerDNP} (see Fig.\ref{Cartoons}b). 
Such transitions require an electron spin-flip, which can be mediated by either the hyperfine interaction or the spin-orbital interaction, denoted in Fig.\ref{Cartoons} by the amplitudes ${v'}$ and ${v}$, respectively. 
The hyperfine process 
is accompanied by a change of the $z$-projection of nuclear spin, $\Delta m=\pm 1$ for an $L$-state of the type $T_\pm$, whereas for the spin-orbital process $\Delta m=0$. 
Without loss of generality, here we focus on the case of transport through a $T_+$ state, pictured in Fig.\ref{Cartoons}.
The resulting coherent dynamics in the combined Hilbert space of electron and nuclear degrees of freedom is thus described by a quantum walk like that shown in Fig.\ref{Cartoons}a.
The topological transition, which is accompanied by the formation of a non-decaying dark state, 
leads to a prediction of threshold-like pumping of nuclear polarization, along with strong suppression of current due to the divergence of dwell time 
at the threshold. 

Because this problem exhibits a topological phase transition, here our main aim is to discuss this interesting behavior from a general point of view; the detailed analysis of implications for spin-blockade is postponed to a separate publication~\cite{SOvsHF}. Here we just note that an ensemble of nuclear spins can be described by a single variable $m$ in the ``giant spin'' approximation of constant hyperfine interaction \cite{Taylor03}, which ignores decoherence 
arising from the more realistic position-dependent interaction \cite{Khaetskii02}. 


The configuration space of the problem is defined by the electron 
states $\Ket{L}$, $\Ket{R}$ and by the total nuclear polarization, taking integer values $-\infty < m < \infty$. 
Thus the Hilbert space has a tensor product structure:
${\rm span}\{\Ket{m}\otimes\Ket{L/R}\}$.
In this basis, illustrated in Fig.\ref{Cartoons}a, the state of the system $\Ket{\psi}$ is described by the amplitudes $\psi_m^L = \Amp{m\, L}{\psi}$ and $\psi_m^R = \Amp{m\, R}{\psi}$, and evolves according to the equations of motion
\begin{eqnarray}\label{EOM1}
\begin{array}{l}
 i\hbar \, \dot{\psi}^L_m\ =\ \varepsilon_L\, \psi^L_m \ + \ {v}\, \psi^R_m \ +\  {v'} \psi^R_{m+1}\\
i\hbar \, \dot{\psi}^R_m\ =\ \tilde{\varepsilon}_R\, \psi^R_m\ +\ {v}\, \psi^L_m\  +\ {v'} \psi^L_{m-1}.
\end{array}
\end{eqnarray}
Without loss of generality, we choose ${v}>0$ and ${v'}>0$.
The on-site energy $\tilde{\varepsilon}_R = \varepsilon_R - i\hbar\gamma/2$ for the $R$-states has an imaginary part that accounts for the decay of these states with rate $\gamma$, while the on-site energy $\varepsilon_L$ is real. 

Now, suppose the system is initialized to the $L$-state
\begin{eqnarray}
  \label{IC} \psi^L_m = \delta_{m,0}, \quad  \psi^R_{m} = 0 
\end{eqnarray}
at time $t = 0$, and allowed to evolve freely under the equations of motion (\ref{EOM1}). 
Because of translational invariance, we can equivalently start anywhere on the 
$L$-sublattice. 
Under the dynamics (\ref{EOM1}), the wavepacket describing the quantum walker spreads throughout the lattice and leaks out through its components on the $R$-sites,
decaying completely as $t\rightarrow\infty$. What is the average displacement achieved by the particle before leaking out? 
More precisely, given the ability to detect the site $m$ from which the decays occurs, and thereby measure the decay probability distribution $P_m$ (see Fig.\ref{Cartoons}c), we would like to find
\begin{eqnarray}
  \label{DeltaM}  \Avg{\Delta m} \equiv \sum_m m\, P_m 
,\quad
P_m=\int_0^\infty \gamma | \psi^R_m(t)|^2 \, dt. 
\end{eqnarray}
Although $\Avg{\Delta m}$ can be obtained from an
explicit calculation involving the system's time evolution operator, here we will pursue a less direct but more rewarding approach that helps to uncover the topological structure behind the solution.
The result is supported by numerical simulations, which also allow us to test various features of the model such as its robustness against decoherence. 

As a first step in the calculation of $\Avg{\Delta m}$, we note that the norm of a quantum state $\Ket{\psi}$ evolves according to $\frac{d}{dt}\Amp{\psi}{\psi} = i\Bra{\psi}(\hat{H}^\dag - \hat{H})\Ket{\psi}$.
For Hermitian systems, $\hat{H}^\dag = \hat{H}$ and $\frac{d}{dt}\Amp{\psi}{\psi} = 0$.
However, our system is non-Hermitian due to the complex energy $\tilde{\varepsilon}_R$, 
and, as seen from the equations of motion (\ref{EOM1}),
decays according to 
$\frac{d}{dt}\Amp{\psi}{\psi} = -\sum_m \gamma \vert \psi^R_m\vert^2.$
The decay is thus described as a sum over {\it local} terms accounting for the decay from each site of the lattice, Eq.(\ref{DeltaM}), with $\sum_m P_m = 1$.

It is beneficial to pass to the momentum representation,
$\psi^R_m = \frac1{2\pi}\oint dk\, e^{i k m} \psi^R_k$, where the integral is taken over the Brillouin zone $-\pi \le k < \pi$.
Due to the translational invariance of the system 
(\ref{EOM1}),
the equations of motion in the Fourier representation break up into $2 \times 2$ blocks, one for each momentum $k$:
\begin{eqnarray}
  \label{EOMk} i \hbar \, \frac{d}{dt}
  \left(
    \begin{array}{c}
      \psi^L_k \\
      \psi^R_k
    \end{array}
  \right)
   =
  \left(
     \begin{array}{cc}
       \varepsilon_L & A_k \\
       A_k^* & \tilde{\varepsilon}_R
     \end{array} \right)
  \left(
    \begin{array}{c}
      \psi^L_k \\
      \psi^R_k
    \end{array}
  \right),
\end{eqnarray}
with $A_k = {v} + {v'} e^{ik}$.
The two-component wave function for each $k$ evolves independently of the others, with $p_k(t) \equiv |\psi_k^L(t)|^2 + |\psi_k^R(t)|^2$, the probability density to find the system with momentum $k$ at time $t$, decaying as $\partial_t \, p_k = -\gamma |\psi_k^R(t)|^2$.

Writing $m$ as a derivative with respect to $k$ via $ m\, \psi_m^R = -\frac{i}{2\pi}\oint dk \, \frac{d}{dk}\left(e^{i k m}\right) \psi^R_k$ and integrating
by parts to move the derivative onto $\psi^R_k$, we 
bring Eq.(\ref{DeltaM}) to the form
\begin{eqnarray}
  \label{DeltaM_k}  \Avg{\Delta m} = i\gamma \int_0^\infty dt \oint \frac{dk}{2\pi}\, {\psi^R_k}^* \, \frac{\partial \psi^R_k}{\partial k}.
\end{eqnarray}
Next, we use the polar decomposition $\psi^R_k(t) = u_k(t) e^{i \theta_k(t)}$, where $u_k = |\psi^R_k(t)|$ and $\theta_k = \arg\{\psi^R_k(t)\}$. We assume that $u_k(t) > 0$ for all $t > 0$, which follows from Eq.(\ref{EOMk}) after some algebra \footnote{An explicit form of the evolution operator, found from Eq.(\ref{EOMk}), 
gives $\psi^R_k(t)$ which is nonzero at all $t>0$ 
except when both $\varepsilon_L = \varepsilon_R$ and $\frac14\gamma < |A_k|$.}. 
Using the fact that $\oint dk\, u_k \partial_k u_k = 0$ is an integral of a total derivative over a closed contour, we 
rewrite Eq.(\ref{DeltaM_k}) as
\begin{eqnarray}
\label{eq:Delta m 2}
 \Avg{\Delta m} &=&  -\gamma\int_0^\infty dt \oint \frac{dk}{2\pi}  |u_k(t)|^2\, \frac{\partial \theta_k}{\partial k} \\
\label{eq:Delta m 3}
&=& \oint \frac{dk}{2\pi} \int_0^\infty  dt \, \frac{\partial p_k}{\partial t}\, \frac{\partial \theta_k}{\partial k}
,
\end{eqnarray}
where we replaced $-\gamma |u_k(t)|^2$ by $\partial_t\, p_k$ in Eq.(\ref{eq:Delta m 2}).
With the help of integration by parts in the integral over $t$, the time derivative can be moved from $p_k$ onto $\partial_k\theta_k$, giving $\Avg{\Delta m} = \mathcal{I}_0 -\int_0^\infty dt \oint \frac{dk}{2\pi}\, p_k \, \partial_t(\partial_k\theta_k)$, with 
\begin{eqnarray}
  \label{BdyTerm} \mathcal{I}_0 = \oint\frac{dk}{2\pi}\ \left[ p_k \left.\frac{\partial\theta_k}{\partial k}\right\vert_{t=0}^{\infty}\right].
\end{eqnarray}
%
We will now show that the boundary term $\mathcal{I}_0$ provides the only non-zero contribution to the integral (\ref{eq:Delta m 3}).
First, we use integration by parts on the integral over $k$ to obtain 
\begin{eqnarray}
\label{eq:int_dk} 
-\int_0^\infty dt \oint dk\, p_k \, \frac{\partial^2\theta_k}{\partial t\partial k} = \int_0^\infty dt \oint dk \frac{\partial p_k}{\partial k} \frac{\partial \theta_k}{\partial t}
.
\end{eqnarray}
%
As demonstrated below, this integral vanishes  
because $p_k$ and $\partial_t\theta_k$ 
are both {\it even} functions of $k$.

In order to see that $p_k$ and $\partial_t\theta_k$ are even functions, it is helpful to view the evolution (\ref{EOMk}) within each $2 \times 2$ $k$-subspace as the precession of a decaying pseudospin in a (complex) magnetic field with $z$-component $\varepsilon_L-\tilde{\varepsilon}_R$
and transverse component of magnitude $2|A_k| = 2|{v} + {v'} e^{ik}|$. 
Because $|A_k| = |A_{-k}|$, a static rotation 
about the $z$-axis maps $\hat{H}_{-k}$ into $\hat{H}_k$, with $\hat{H}_k$ the $2\times 2$ matrix in Eq.(\ref{EOMk}):
\begin{eqnarray}
\label{eq:rotated frame}
  e^{-i\phi_k\hat{\sigma}^z}\hat{H}_{-k}\,e^{i\phi_k\hat{\sigma}^z} = \hat{H}_k
,\quad
\phi_k = \arg\{A_k\}
.
\end{eqnarray}
Given that the initial state (\ref{IC}) is oriented along the $z$-axis in pseudospin-space for all $k$, 
in the rotated frame (\ref{eq:rotated frame}) the pseudospin associated with the momentum $-k$ performs the {\it identical} evolution to that of the pseudospin associated with momentum $k$. Because the state (\ref{IC})
has equal magnitude in all momentum sectors,
the moduli of the $k$ and $-k$ pseudospins are equal for all times, $p_k(t) = p_{-k}(t)$.
Furthermore, from (\ref{eq:rotated frame}) their phase difference is time-independent,
$\theta_{-k} = \theta_k - 2\phi_k$,
which proves the claim. 

To evaluate $\mathcal{I}_0$, we use the facts that all $k$-states are initially occupied with equal probability $p_k(t=0) = 1$, and that the state decays completely, $p_k(t\rightarrow\infty) = 0$.
Substituting these values in Eq.(\ref{BdyTerm}), we find
\begin{eqnarray}
  \label{winding}  
\Avg{\Delta m} = - \oint \frac{dk}{2\pi} \, \frac{\partial\theta^0_k}{\partial k}
, \quad 
\theta^0_k \equiv \lim_{t\rightarrow0^+}\theta_k(t)
.
\end{eqnarray}
Although $u_k(0) = 0$, the limit $t\to 0^+$ is well-defined.

Expression (\ref{winding}) is a surprising result: the expected displacement of the particle as it spreads out and decays 
is equal to the {\it winding number} of the relative phase between components of the Bloch wave function.
In particular, this means that $\Avg{\Delta m}$ can only take on {\it integer} values.  
Using $\psi_k^R(dt) = -i A_k^* dt/\hbar$, we have
$\theta_k^0 = \arg\{-iA_k^*\}$. 
It is thus immediately clear that there are two possible situations depending on
whether or not $A_k=v+v'e^{ik}$ wraps the origin as $k$ is taken around the Brillouin zone: $\Avg{\Delta m}=1\,(0)$ when ${v'} > {v}$ (${v} > {v'}$).

\begin{figure}
\includegraphics[width=3.4in]{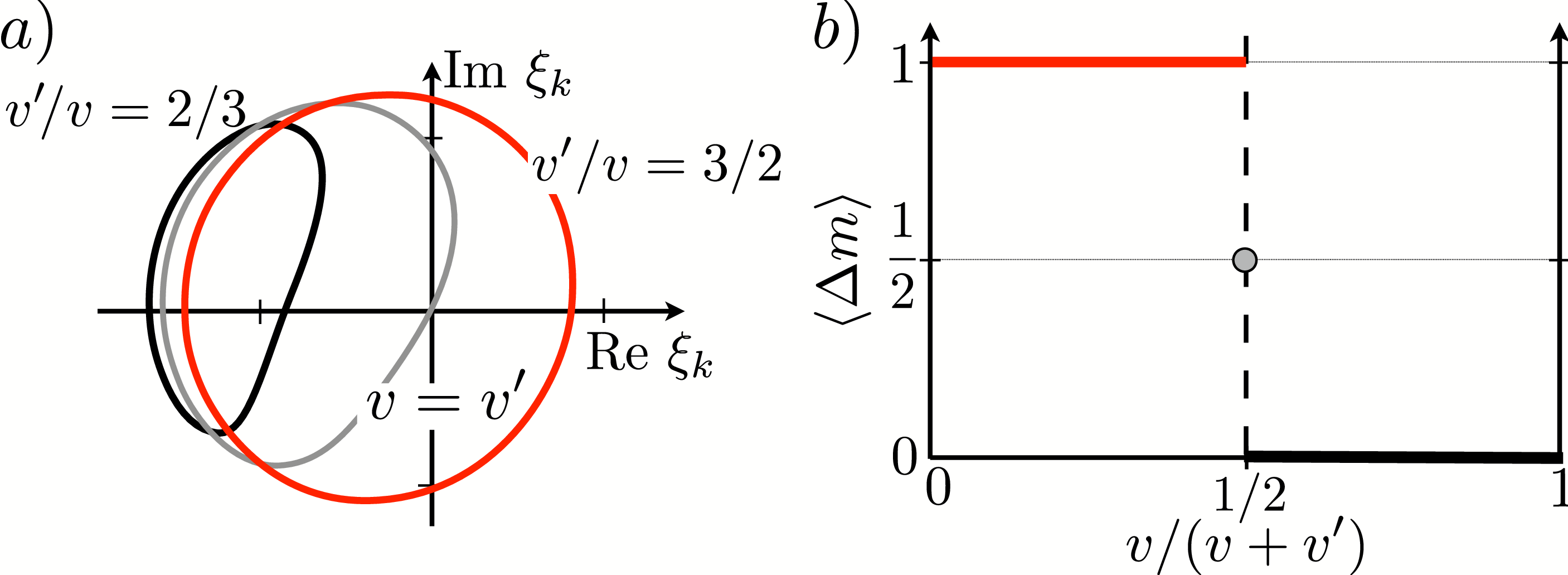}
 \caption[]{a) The winding number change for an eigenstate of the Hamiltonian (\ref{EOMk}). Here we plot the component ratio $\xi_k=\psi^R_k/\psi^L_k$ {\it vs.}  momentum $k$ for $-\pi<k<\pi$.
b) The expected displacement, Eq.(\ref{DeltaM}),
after full decay of the initially localized state.
The quantization of $\langle \Delta m \rangle$ is topological in nature, and is linked to the winding of $A_k = {v} + {v'} e^{i k}$ around the origin. }
\label{Analytics}
\vspace{-4mm}
\end{figure}

It is perhaps not entirely obvious from this discussion that the transition at $v=v'$ is a characteristic of the Hamiltonian rather than of the initial state. 
To clarify this point, we examine the eigenstates of 
$\hat H_k$ and plot the ratio of their components $\xi_k=\psi^R_k/\psi^L_k$ in the complex plane. As 
shown in Fig.\ref{Analytics}a, the winding number about the origin changes from 1 at $v'>v$ to 0 at $v'<v$.

Furthermore, one of the eigenvalues of $\hat H_k$ becomes real at the transition $v=v'$, because $A_k$ vanishes for $k=\pi$. This indicates the formation of a non-decaying {\it dark state} with $\psi_{k=\pi}^R=0$; under these conditions,
the $k=\pi$ component of the initial state (\ref{IC}) remains stuck on the $\Ket{L}$ sublattice for all $t$. 
Dark states formed in the nuclear subspace, as found here, 
can arise as fixed points of cooling processes~\cite{Taylor03}. 

As pointed out in Ref.\cite{Raikh94} (see also \cite{Brandes05}), dark states 
in quantum dots can result in current suppression due to the Dicke effect. 
In our case, the average decay time 
\begin{eqnarray}
\label{eq:bar tau}
  \bar{\tau} = -\int_0^\infty t\, \frac{d}{dt}\Amp{\psi}{\psi} \, dt 
= \oint \frac{dk}{2\pi}\int_0^\infty  p_k(t)\, dt
\end{eqnarray}
%
may become very long near the transition (here we used $\frac{d}{dt}\Amp{\psi}{\psi} = \frac1{2\pi} \oint dk \, \partial_t p_k$ and integrated by parts).
Close to the transition ${v} = {v'}$, when 
$|A_{k \approx \pi}|\ll \hbar\gamma$, 
the dynamics (\ref{EOMk}) yields $p_k(t) \approx \exp(-\Gamma_k t)$, where
$\Gamma_k$ is given by Fermi's Golden Rule:
  $\Gamma_k = |A_k|^2\gamma/[(\varepsilon_L-\varepsilon_R)^2 + (\hbar\gamma/2)^2].$

Substituting these expressions into Eq.(\ref{eq:bar tau}), and using the change of variables $z = e^{ik}$, we get
\begin{eqnarray}
  \bar{\tau} = \frac{(\varepsilon_L-\varepsilon_R)^2 + (\hbar\gamma/2)^2}{2\pi i\gamma}\, \oint \frac{dz}{({v}z + {v'})({v} + {v'}z)},
\end{eqnarray} 
where the integral is taken over the unit circle $|z| = 1$.
Using the residue theorem, we see that the decay time indeed diverges at ${v'}={v}$ as $\bar{\tau} \propto 1/|{v}-{v'}|$. 

Conspicuously, neither the quantization of $\Avg{\Delta m}$, nor the discontinuity at ${v} = {v'}$, seem to depend on the values of 
the decay rate $\gamma$ or the energies $\varepsilon_{L/R}$.
Furthermore, the analysis leading up to Eq.(\ref{winding}) 
goes through even if $\gamma$ and $\varepsilon_{L/R}$ are made time-dependent. 
In particular, the integral (\ref{eq:int_dk}) still vanishes because the states $k$ and $-k$ see identical time dependent effective fields, up to a rotation (\ref{eq:rotated frame}). 
This suggests, among other things, that the sharp transition shown in Fig.\ref{Analytics}b 
survives dephasing due to classical noise on the energy levels $\varepsilon_L$ and $\varepsilon_R$.

\begin{figure}
\includegraphics[width=3.0in]{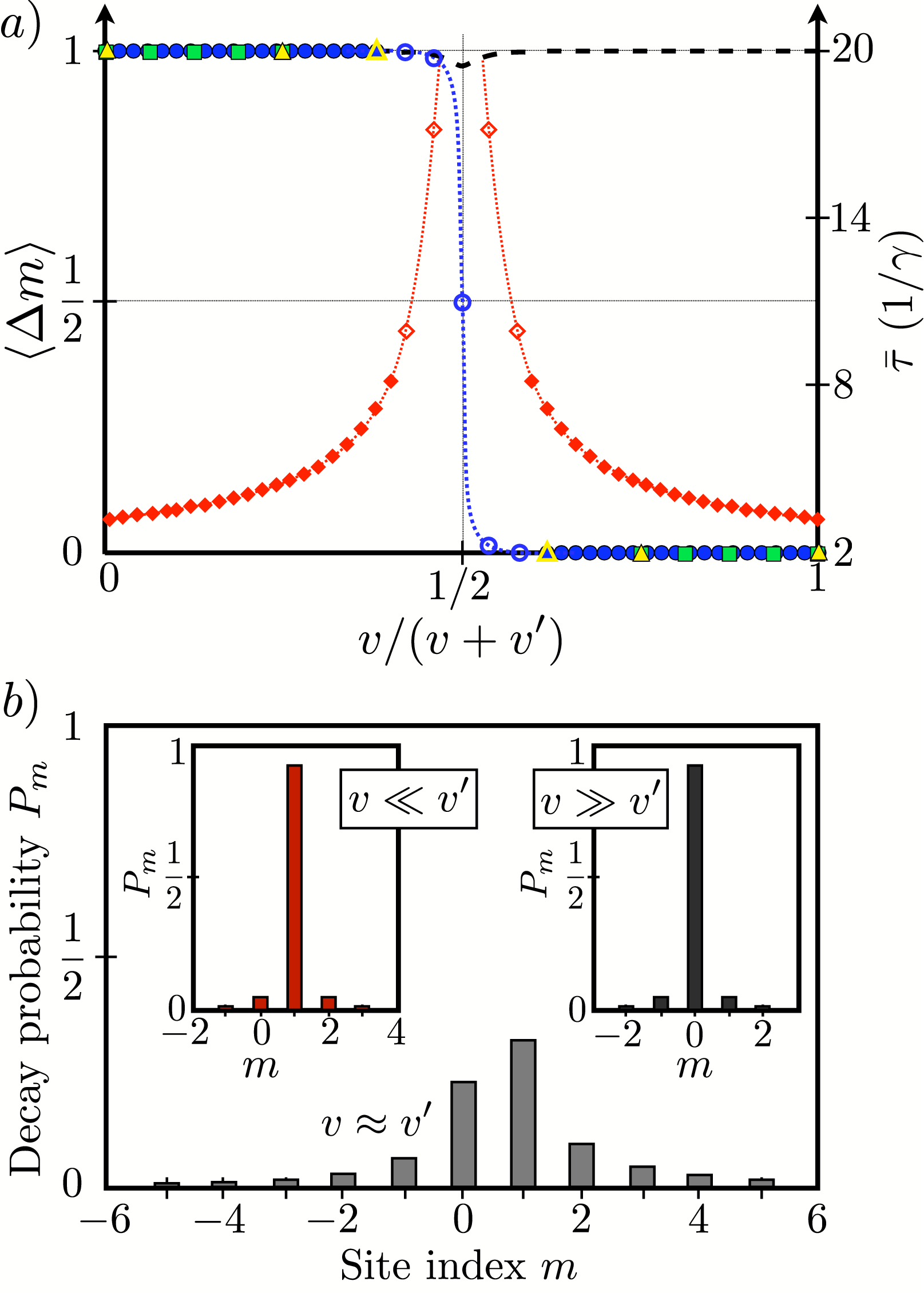}
 \caption[]{Results of simulation with finite chain of length $N = 51$ unit cells and $\gamma = \varepsilon_L-\varepsilon_R = 1$.
a)  Displacement $\Avg{\Delta m}$ (blue circles) and decay time $\bar{\tau}$ (red diamonds). 
Filled symbols were obtained by evolving the wave function up to time $T = 100$.  
Decay slows down near the critical point, where longer running time is required (open symbols, $T = 500$).
The black dashed line shows $1 - |\psi(T)|^2$, used to monitor completion of the simulation.
Dotted lines are smooth spline fits added as an aid to the eye.
Incomplete decay due to finite running time $T$ and finite length $N$ appear as rounding of the step. Results of simulation with $\Gamma_2 = 10$ for linear damping (green boxes) and repeated projective measurement (yellow triangles) show that quantization survives $L-R$ decoherence.
b) Decay probabilities $\{P_m\}$ at ${v}/{v'} = \frac19,\,0.85,\,9$.
The distribution becomes broad near the transition $v = v'$, while the mean remains quantized.
}
\label{Simulation}
\vspace{-4mm}
\end{figure}

To investigate this remarkable indifference to dephasing, we have performed direct numerical simulations of the equations of motion (\ref{EOM1}) up to a fixed time $T$ and restricted to a finite chain of 51 unit cells.
During each time step $t_n < t < t_n + \Delta t$, we evolve the state forward in time and bin the probability of decay from each unit cell.
In Fig.\ref{Simulation}a we show the results for the mean displacement (\ref{winding})
and the decay time $\bar{\tau}$, obtained using the distribution $P_m$ (see Fig.\ref{Simulation}b), and the formula $\bar{\tau} = \sum_n |\psi(t_n)|^2 \Delta t$.

This simulation, showing clear quantization, was then altered to investigate the robustness against decoherence. We modified the simulation to evolve the system's density matrix, adding an exponential damping of the $L-R$-off-diagonal elements with rate $\Gamma_2$.
Due to the increased  time and memory requirements, we could only simulate smaller systems over a more sparse sampling of points.
However, the results do show a relatively well-formed step (green boxes), consistent with expectation.
Similarly, the step appears to be robust against a stronger form of decoherence where the density matrix is repeatedly projected onto the $L$ and $R$ subspaces at a fixed time interval $\Delta\tau = 1/\Gamma_2$ (yellow triangles).

The transition is not robust against {\it all} types of noise, however.
Any variations of the amplitudes ${v}$ and ${v'}$ in time
 will in general broaden and distort the step. Likewise, we do not expect the sharp step to survive perturbations that break translational symmetry.


In our motivating example of spin-blockaded transport in double quantum dots, the topological transition would be manifested as an abrupt change in the nuclear spin pumping rate as the relative strengths of spin-orbit and hyperfine matrix elements are varied, for example, by tuning gate voltages to change the electrostatic potential felt by the electron.
Because pumping thresholds in various parameters are ubiquitous in such systems, to unambiguously identify a pumping threshold with the topological transition discussed here one must correlate the appearance/disappearance of nuclear spin pumping with 
a {\it decrease} in the current through the system resulting from the diverging dwell time $\bar \tau$, which is not expected for pumping thresholds of other origins.

We thank B. I. Halperin and J. Krich for helpful discussions, and acknowledge 
support from W. M. Keck
Foundation Center for Extreme Quantum Information
Theory, from 
a National Science Foundation Graduate Research Fellowship (M.R.), and from
the NSF Grant No. PHY05-51164 (L. L.). 


\end{document}